\documentclass{sig-alternate}

\usepackage{epstopdf}
\usepackage{hyperref}
 \hypersetup{ 
    colorlinks,%
    citecolor=black,%
    filecolor=black,%
    linkcolor=black,%
    urlcolor=black 
  } 
\usepackage{graphicx,algorithm,algorithmic}
\usepackage{amsmath}

\usepackage{graphicx,type1cm,eso-pic,color}

\def\sharedaffiliation{%
\end{tabular}
\begin{tabular}{c}}

\begin{document}

\conferenceinfo{\bigskip ModelsGamification12,}{October 01 - 05 2012, Innsbruck, Austria}
\CopyrightYear{2012} 
\copyrightetc{Copyright 2012}

\title{The Art of Software Design, a Video Game for Learning Software Design Principles}
\subtitle{Gamification of Software Design Learning Objectives}
%
%
%
%
%

\numberofauthors{3} 
%
\author{
%
%
\alignauthor
Dave R. Stikkolorum\\
\email{drstikko@liacs.nl}    
\alignauthor
Michel R.V. Chaudron\\
\email{chaudron@liacs.nl}    
\alignauthor 
Oswald de Bruin\\
\email{snake.ossi@gmail.com}    
\and
\\
\sharedaffiliation
\affaddr{Institute of Advanced Computer Science, Leiden University}\\
       \affaddr{Niels Bohrweg 1, Leiden, the Netherlands}\\       
\\
}

\maketitle
\begin{abstract}
This paper introduces our gamification of a part of our software design curriculum. Based on typical design principles a motivating learning game is developed to train students in software design. We use Bloom's taxonomy to determine learning objectives. We keep the player engaged with direct feedback in a challenging level based game with increasing complexity. Players can evaluate their design actions with the help of the visualisation of control and data flows. The main learning objective: applying design principles, fits the game's main activity. This supports the learning by doing approach of lecturers. A user test indicates possible learning effects and a playable game.
\end{abstract}
\category{D.2.2}{SOFTWARE ENGINEERING}{Design Tools and Techniques}[Design Principles]
\category{D.2.10}{SOFTWARE ENGINEERING}{Design}
\category{K.3.1}{COMPUTERS AND EDUCATION}{Computer Uses in Education}[Game Based Learning]

\terms{Design, Didactics, Gaming}

\keywords{Gamification, Video Game, Education, Software Design, Modelling} 

\section{Introduction}

Designing of software systems is one of the difficult tasks in the field of software engineering. It is no surprise that learning software design is a difficult task for students too. They have to face the challenge of abstracting structure and behaviour for possible software solutions. Most lecturers choose the approach of letting students practice with artificial cases and repeat this a couple of times. We can not expect that this approach always motivates students. In this paper we introduce our gamification of a part of our software engineering curriculum. We aim to motivate students to learn software design by providing them learning material in the form of an interactive computer game.

In the field of software engineering a couple of game environments were introduced with programs like BlueJ\footnote{\url{http://www.bluej.org}} and Alice\footnote{\url{http://www.alice.org}} to support students in learning to program and understand the concept of object orientation. However in this type of programs only a game world is provided, you have to create the game yourself. In the field of architecture modelling Groenewegen et al. \cite{Groenewegen_2012} introduced a game for validating architecture models.  Groenewegen et al. used a board game where the actual models play a role. We also aimed to create a game where the models itself are game elements. We used a video game. As far as we know no such game in the field of software design exists. 

In this paper we discuss our gamification approach, give insight in the game itself and discuss early findings based on user tests. In section \ref{method} we describe our method, in section \ref{game_design} we describe the game design. After a `walk through' of the game in section \ref{walk_trough} we evaluate and discuss in sections \ref{eval} and \ref{discussion}. Finally we conclude and propose future work in section \ref{conclusion}.

\section{method}\label{method}
The aim of this study is to create a playable game to explore the possible support for the learning of students. It is created using an iterative approach. The authors came together periodically and discussed and improved their different ideas until there was a first version.

The learning objectives of the game are chosen using\\ Bloom's taxonomy \cite{Krathwohl_2002} combined with a set of general design principles \cite{Martin_2000} : coupling, cohesion, information hiding and modularity. Bloom's taxonomy is widely used by lecturers. `Design principles' is a typical software design subject.

For further improvement we tested the game with a user test in combination with a simple version of the `think aloud' \cite{lewis1982using} method. We simply asked the test users to tell us their thoughts while making steps in the game. We know this is no complete validation. As mentioned before there is no game to compare with. To demonstrate the learning effect of the game we see a challenge in future research. 

\section{game design}\label{game_design}
In this section we discuss the learning objectives, the type of the game, game levels, the role of UML, the game mechanics and the software that was used to implement the game.

\subsection{Learning Objectives, The Aim of The Game}
The main learning objective of the game is to understand and be able to apply main software design principles: coupling, cohesion, information hiding and modularity.

In this paper we use the term `design' rather than `modelling', because we see modelling as a vehicle for designing. The game uses modelling as a vehicle for:
 i. explaining software design principles. ii. the usefulness of models for abstract reasoning about a design. iii. finding the proper abstractions for representing a system in a domain.

\subsection{Type of Game}
We chose to make the game a puzzle game. The activity of looking for a solution that for example provides balance between coupling and cohesion is a very similar task in comparison to solving a puzzle. 

\subsection{Game Levels}
For each of the design principles, there is a set of levels that offers puzzles of increasing complexity. A puzzle offers a design fragment - typically a set of classes, attributes and some methods - and asks the player to complete the design. The moves that a player can make differ per level. For simple levels, there are predefined classes and methods that the player can move around and connect to existing classes in the design. At more advanced levels, the player can also create new classes. The initial levels of the game test for understanding the main concepts in isolation. Subsequent levels offer puzzles that require combined understanding of multiple design principles.

\subsection{UML}
Although UML is used for representing the designs, the game is not intended for learning the syntax of UML. We used a very simple subset of UML and tried to stay away from Object-orientation if the same principle could be explained in a more general manner. 

\subsection{Game Mechanics}
There are several mechanics that we used to achieve an educational and engaging game. In most of the cases the mechanics mentioned below affect both educational and engagement goals.

\subsubsection{direct, visual and audio feedback}
The game offers feedback through two means: i. each level is scored through an evaluation mechanism supported by audio and visuals. ii. the user is given feedback through visualization of control flow and data flow. When a player moves or modifies elements feedback is provided directly and the score adjusts. On every logical action a sound is played. 

\begin{figure}
\centering
\includegraphics[scale=0.42]{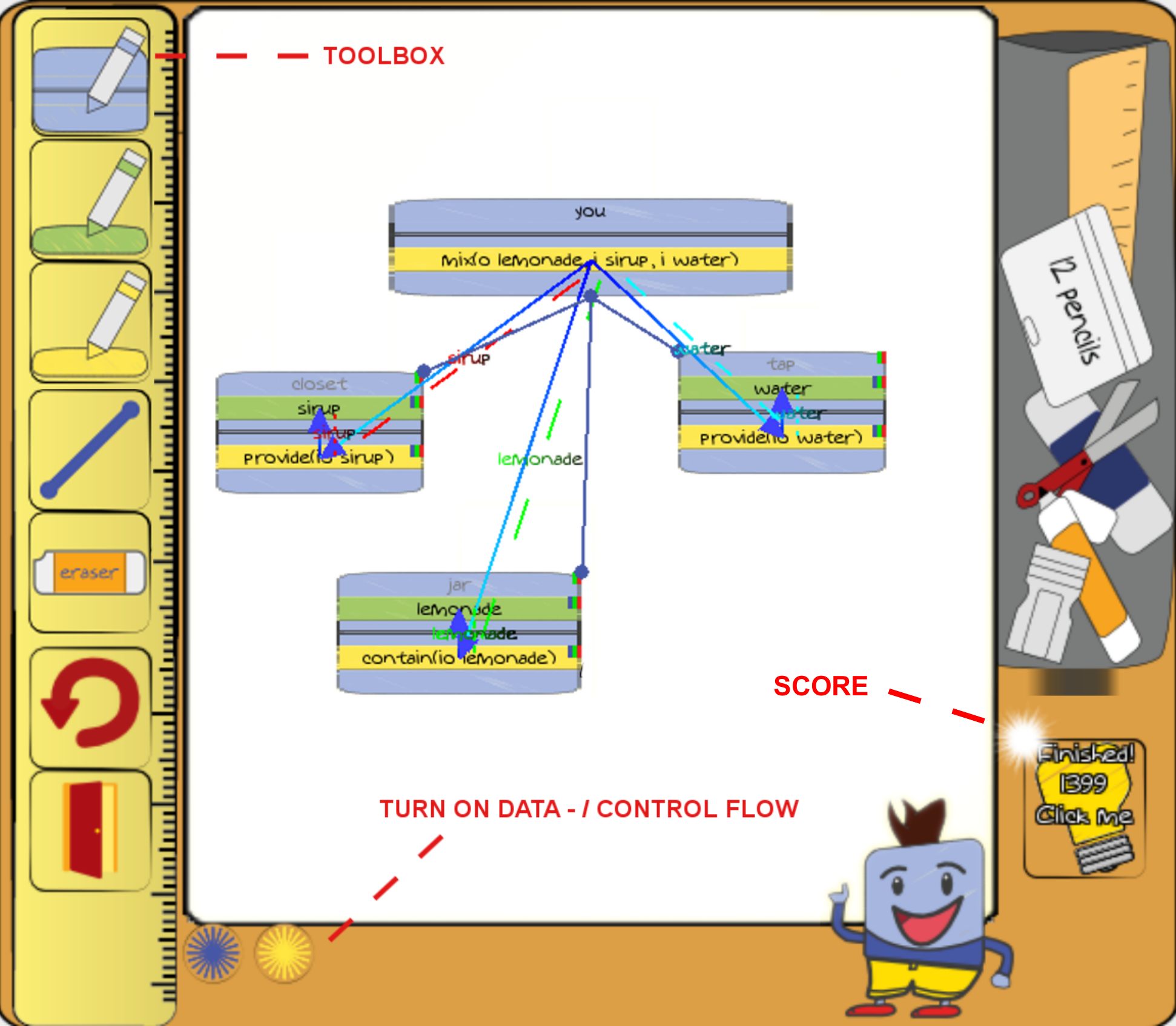}
\caption{Data and control flows. Control flows are represented by an arrow. Data flows are dashed. Comments are made in red.}\label{fig:data-control-flow}
\end{figure}

To support a player in decision making, the data and control flow of a puzzle can be visualized. This can be seen in figure \ref{fig:data-control-flow}. In this way a player can check the effect of placing an attribute or method in a certain class in comparison with the assignment of the puzzle. We think this is a valuable element of the game from the perspective of self evaluation.

\subsubsection{level unlocking}
To experience achievement, not all puzzles are playable from the moment the games starts. They have to be unlocked by finishing other puzzles before being able to play that particular one. In this way it is not possible to skip levels. A player needs to learn (by doing) a couple of subjects before they can work on dependent puzzles. By applying the unlock mechanism we aim to create awareness of the dependencies between the learning objectives and the concepts.

\subsubsection{choice of path}
As mentioned in other research \cite{Moser_1997} \cite{Przybylski_Rigby_Ryan_2010} a user is more engaged if the freedom of choice increases. We designed the levels in a way that one can start from different starting points. Even when a player gets stuck he/she is always able to return and take another path.  

\subsubsection{multiple solutions}
Due to the context of the game, software design, the puzzles do not have one best solution. If a design respects the design principles there still can be a couple of different solutions. A design can be better than another considering the problem domain or other contextual factors. We preserved this real-life situation and made this also occur in the game.

\subsection{Scoring Metrics}
Every puzzle is scored by an evaluation script. This script checks if the design that was created matches possible solutions based on the design principles.

\subsubsection{coupling}
To determine coupling within a puzzle we used a simple approach. We only used the CBO (coupling between object classes)\cite{Chidamber_Kemerer_1994} metric.

\textit{CBO per class is: the number of other classes it connects to}

Per design the average would be the sum of all the CBO (indexed by \textit{i}) divided by the total of classes (n): \\
$average CBO =\frac{\sum\limits^{n}_{i=1} CBO_i}{n}$

\subsubsection{cohesion}
\begin{figure}
\centering
\includegraphics[scale=0.52]{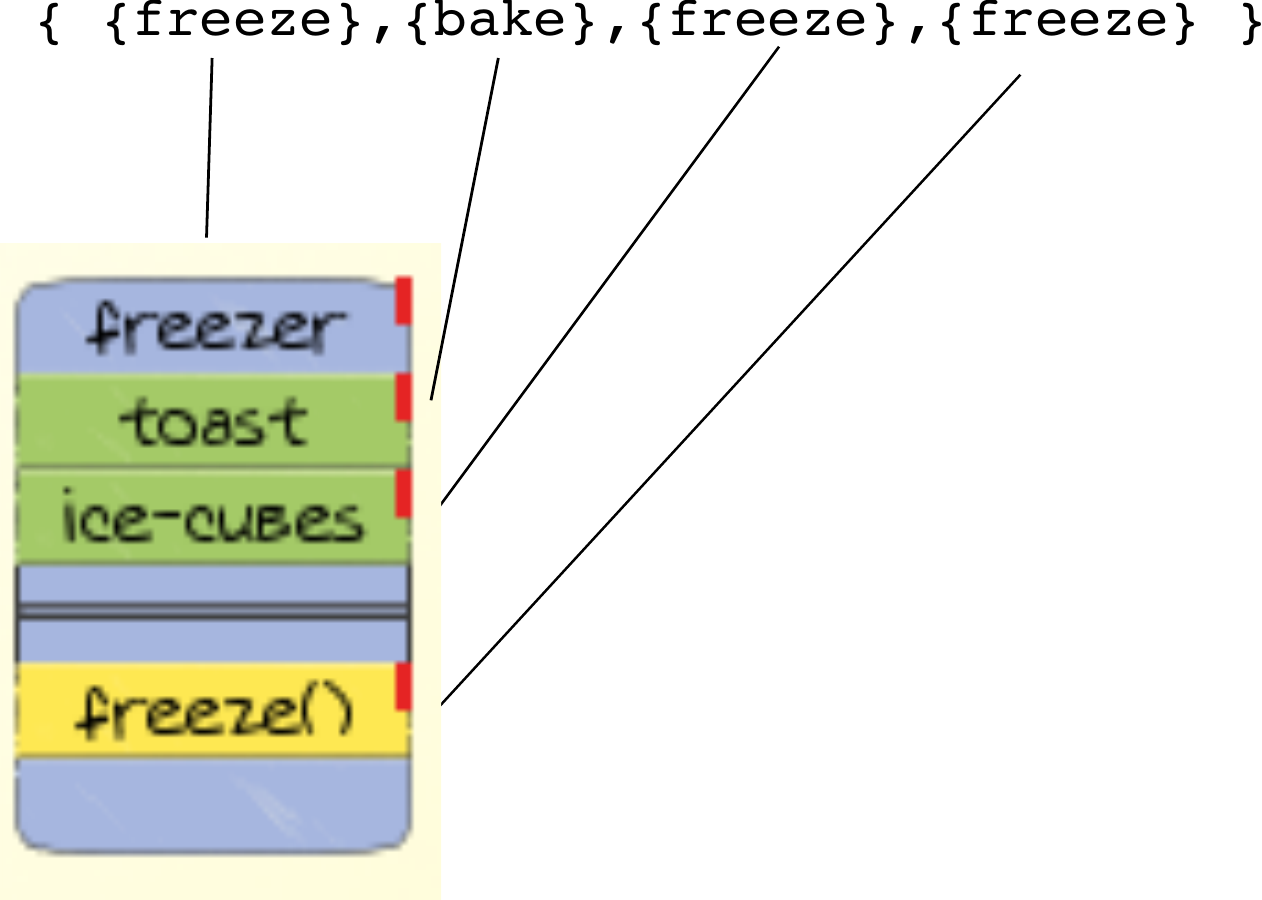}
\caption{Connected keywords to classes, attributes and methods}\label{fig:cohesion_explained}
\end{figure}

Every class, attribute and method has one or more keywords attached to it. We use these keywords to determine if two elements are related, if they are cohesive. This is illustrated in figure \ref{fig:cohesion_explained} .

The cohesion evaluation script compares if all items of a class (including the class itself) have similar keywords. In the example provided in figure \ref{fig:cohesion_explained} only one keyword is attached per element. This can also be two or more. 

Cohesion of a class (CC) is determined as follows:
i. per comparison (indexed by \textit{i}), divide the number of keyword matches (\textit{m}) with all the other elements by the number of keywords (\textit{k}) that are being considered in the comparison. This gives us the match ratio per comparison. ii. divide the sum of these ratio's by the number of comparisons (\textit{n}). This gives us the total index of the similarities of a class, the cohesion:\\
$ CC =\frac{\sum\limits^{n}_{i=1}{\frac{m_i}{k_i}}}{n}$

\subsubsection{information hiding and modularity}
To evaluate the application of information hiding and modularity we used general design patterns. The player has only a limit of choices and is guided to a solution that uses this patterns. The evaluation script checks if the user applied the elements (classes, attributes, methods) in a way that matches the pattern.

\subsection{Software Platform}
The game is constructed with Gamemaker 8.1 Standard\footnote{\url{http://yoyogames.com/gamemaker}}. We made this choice so we could rapidly make fully functional prototypes. Gamemaker has a readily available engine with graphics, mouse events, scripting and other features that can be used in games.\vfill\eject

\section{the game}\label{walk_trough}
In this section we provide an overview of the flow of a game session.
The main aim of the game is to complete every puzzle.

`The Art of Software Design' \footnote{The Art of Software Design : \url{http://aosd.host22.com}}$^{,}$\footnote{Trailer : \url{http://www.youtube.com/watch?v=xn1E2dU-\_zg}} starts with an opening screen and gives the player the opportunity to personalise his game by entering his name. This name is used as a saved game and by clicking it, it will resume after the last finished puzzle. After entering the welcome screen a puzzle-tree appears. This tree consists of puzzles based on software design principles or combination of those principles. Players have to unlock upper level puzzles before they can do the next puzzles. 

\begin{figure}
\centering
\includegraphics[scale=0.5]{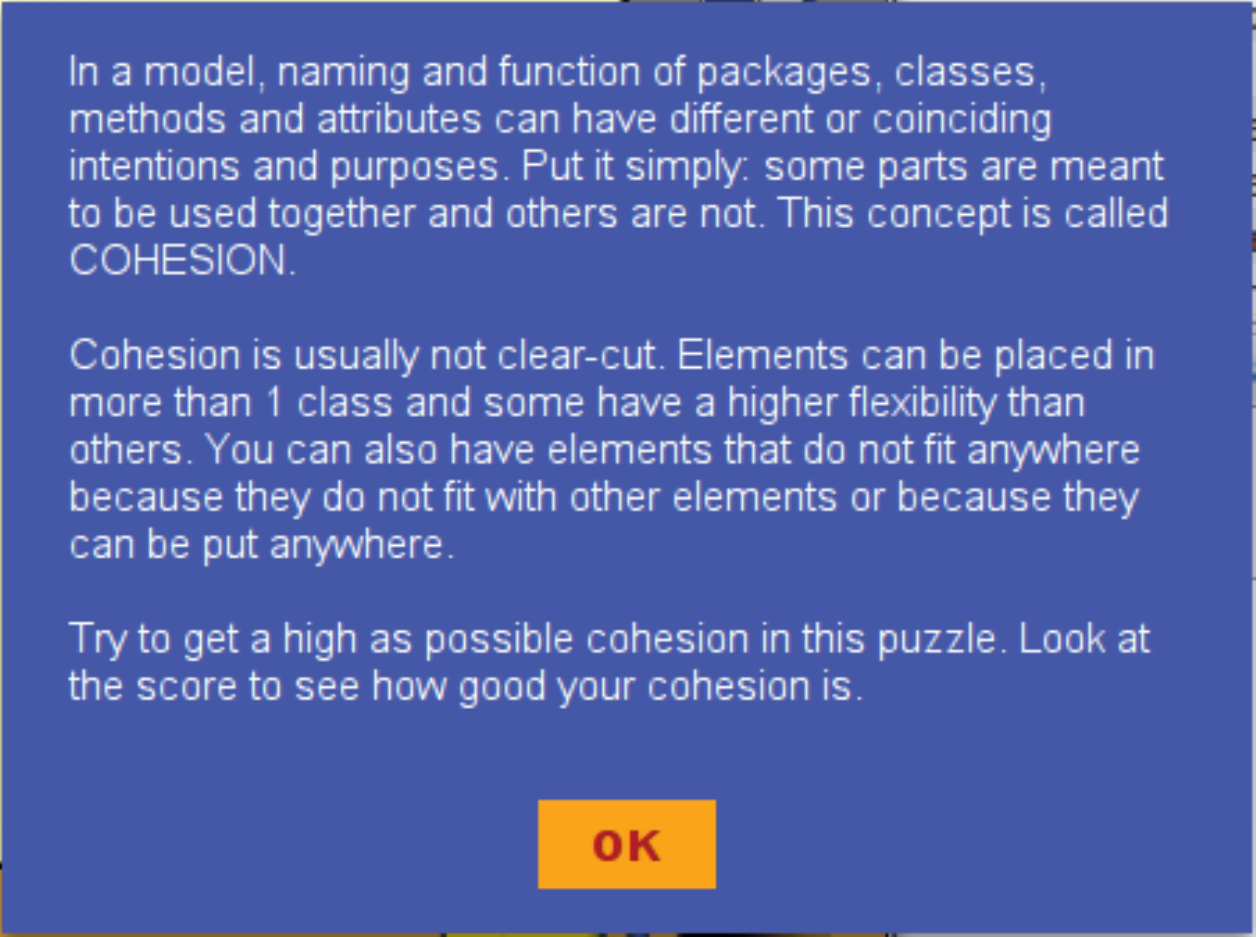}
\caption{Pop-up with assignment for the cohesion puzzle}\label{fig:cohesion_puzzle}
\end{figure}

Inside a puzzle a welcome messages is showed (fig. \ref{fig:cohesion_puzzle}). This message contains the assignment of the puzzle.
\begin{figure*}
\centering
\includegraphics[scale=0.57]{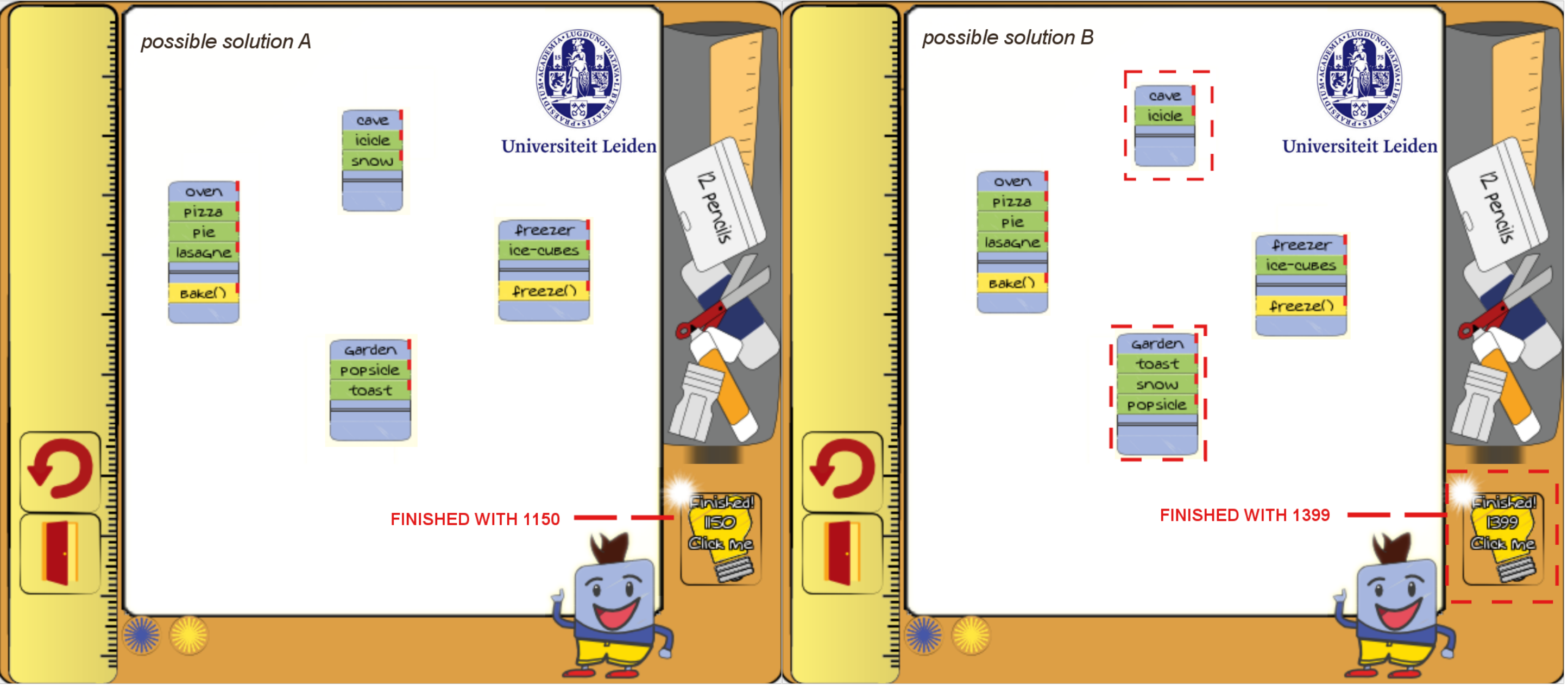}
\caption{2 possible solutions for the cohesion puzzle. The red striped boxes show the differences.}\label{fig:cohesion_puzzle_sol}
\end{figure*}
Typical tasks a player has to fulfil are such as placing attributes and operations in the right classes (responsibility driven approach), connect the right classes (associate) given a typical design principle. A toolbox is present for adding these items when needed.  A progress indicator shows how close the player is to the solution of the puzzle. As mentioned before the game offers the opportunity to complete a puzzle with different solutions. One can be better than the other. This is shown by the players' score as shown in figure \ref{fig:cohesion_puzzle_sol}.

After completing the puzzle, the player returns to the main screen where certain puzzles are unlocked. From there new challenges are possible.

The game ends when all the puzzles about basic design principles have been fulfilled. After that, in future versions of the game, series of more complex puzzles can be offered.

\section{evaluation}\label{eval}
In a thinking aloud setting we observed the gaming experience of a group  of users and noted their comments.

We noticed that giving freedom of choice was of value. When people got stuck on a puzzle, they tried out different solutions or tried another puzzle and returned to the more difficult puzzle later on.

Players paid more attention to the instructions prior to a puzzle when there were 3 paragraphs of 4 lines of text at maximum. The text was best understood if it had an introductory paragraph, an explanatory paragraph and an assigning paragraph in that order. Some puzzles seem too simple. They were solved without reading the instructions. 

For us it was satisfying to see, that after a while subjects started talking about the puzzles in terms of  `classes, methods and associations', instead of `boxes, blocks and lines', which seems to indicate some unconscious learning.

\section{Discussion}\label{discussion}
Determining the average coupling of the design could be a too rough measure. The average coupling can turn out to be relatively low, while a certain class in the design can have a very high coupling. It may be wise to indicate the user that the coupling of a certain class is too high. 

Although we find it very plausible that we choose to simulate associative thinking with keywords in order to determine cohesion, we have no data that validates that approach. It may be wise to explore correlation between metrics such as (L)COM \cite{Harrison_Counsell_Nithi_1998} to validate our keywords method.

We acknowledge that further study is needed to demonstrate the learning effect, but we think the unconscious learning of concepts like classes, attributes and methods at least indicates a certain learning effect. 

\section{Conclusion}\label{conclusion}
The aim to create a playable learning game seems to be accomplished. Although we are not able to prove that our approach is completely valid, we have some good indicators for future research. A deeper study of the validity of the score metrics is needed. Further research is needed to demonstrate the learning effect of the game. We suggest to study both validation and learning effect in a case study that uses `The Art of Software Design.'
\bibliographystyle{abbrv}
\bibliography{AOSD_a_Video_Game_for_Learning_Software_Design_Principles}

\end{document}